\documentstyle[spie]{article} 
\input{psfig}   

\title{Development of Gold Contacted Flip-chip Detectors with IMARAD CZT} 

\author{T. Narita\supit{1}, P.F. Bloser\supit{1},
J.E. Grindlay\supit{1}, J.A. Jenkins \supit{1}
\skiplinehalf 
\supit{1}Harvard-Smithsonian Center for Astrophysics, 60 Garden St., Cambridge, MA 02138 
}

\authorinfo{Further author information: (Send correspondence to
T. Narita)\\ tnarita@cfa.harvard.edu}

\begin{document} 
  \maketitle 

\begin{abstract}
We present initial results from our evaluation of a gold-contacted
pixellated detector using cadmium zinc telluride substrate produced by
IMARAD Imaging Systems. The Horizontal Bridgman (HB) grown crystals
from IMARAD have been shown to produce high resolution photopeaks, but
they are also seen to have large leakage current.  Our previous tests
with IMARAD CZT showed that the use of indium anodes and gold cathode
improved the resistivity compared to the standard indium-contacted
detectors.  We seek to test whether simple evaporated gold contacts
alone could also reduce the leakage current and thus improve the
spectral resolution, especially in the 10-100 keV energy range.  We
have fabricated several metal-semiconductor-metal (MSM) detectors with
a $4\times4$ array of pixels on $10\times10$ mm substrates.
Measurements of the detectors' leakage current, spectral response, and
temperature sensitivity are presented and compared to IMARAD's ohmic
contact detector and gold contact MSM detectors made of High Pressure
Bridgman (HPB) material. Finally, we show preliminary results from a
tiled flip-chip pixellated detector made using the IMARAD detectors.
\end{abstract}



\section{INTRODUCTION}
\label{sect:intro}  

Due to the considerable potential shown by Cadmium Zinc Telluride
(CZT) in the past several years, the next generation of astronomical
high energy X-ray satellites (Swift, Constellation-X, EXIST) will
almost certainly employ CZT as their hard X-ray detectors.  The
advantages of CZT are numerous: operation at room temperature due to
its large bandgap, effective stopping power at reasonable thickness,
and the lack of polarization effects typically found in CdTe.  A
variety of electrode designs have also been developed to compensate
for the poor hole mobility, and one can now obtain much better energy
resolution from CZT than from scintillators.  These favorable material
characteristics and the new readout designs have made CZT a promising
semiconductor for medical and nuclear technologies, in addition to
astronomy. For a mission such as EXIST\cite{grindlay99}, where the
detector area approaches ten square meters, major issue is the
increase in yield of large volume spectroscopic grade material.
Currently the High Pressure Bridgman (HPB) growth process produces
high resistivity (10$^{11}$ ohm-cm) material which is relatively free
of structural defects.  However, the yield of HPB crystals with
dimensions greater than $10\times10$ mm$^2$ and free of macroscopic
defects is low.  Fabrication of large area detectors will require
tiling many smaller modules and thus the availability of large volume
crystals is essential.

IMARAD Imaging Systems produces CZT which is grown using a modified
Horizontal Bridgman (HB) process\cite{cheuvart90}. This method
produces large area crystals ($40\times40$ mm$^2$) which are 
relatively free of defects.  Pixellated detectors made with IMARAD
material have been characterized and found to have uniform leakage
current and spectroscopic response across
pixels\cite{schlesinger98,narita99}.  By using larger crystals, the
problems associated 
with tiling detector modules into a large area array are greatly
simplified. However, one drawback of the HB process is that the
crystals are typically lower in resistivity than those made with the HPB
process.  A pixellated IMARAD detector which uses indium for the
electrodes has a typical resistivity of about $5\times10^{9}$
ohm-cm\cite{uri99}.  While low resistivity is not critical for
spectroscopy at X-ray energies greater than 100 keV, the large leakage
current significantly degrades the energy resolution at 5-60 keV.

One method to decrease the leakage current of a detector, and thus
improve the
spectral resolution, is the addition of blocking contacts.  Recent
tests using gold cathode and either CdS or In anodes on IMARAD
material showed an increase in the reverse bias resistivity
\cite{narita99}. We speculated that the gold electrode acted as a blocking
contact when used with IMARAD material, and thus it would be a simple
process to improve the detector performance. In this paper, we report
our preliminary results from gold pixellated detectors made with
IMARAD and eV Products CZT.  We compared the resistivity, spectral
response, and the low temperature performance of these detectors.  We
also present results from our tiled flip-chip detector.

\section{DETECTOR DEVELOPMENT} 
\label{sect:CZT_PIN}  

\subsection{Detectors} 
\label{sect:detectors}  

We fabricated two gold contacted detectors using $10\times10\times5$
mm$^3$ CZT samples from IMARAD (hereafter called IMARAD-Au) and two
gold contacted detectors of the same dimension with HPB CZT from eV
Products (hereafter called eV Products-Au). The detectors were
processed at RMD in Watertown, MA.  To insure uniformity, one IMARAD
crystal and one eV Products crystal were each placed side by side in
the evaporation chamber during fabrication.  Evaporated gold was used
for the electrodes, and a shadow mask was used for the pixel
definition.

\begin{figure}
\begin{center}
\begin{tabular}{c}
\psfig{figure=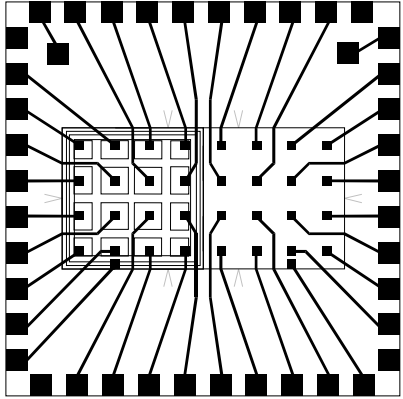,height=6cm} 
\end{tabular}
\end{center}
\caption[detectors] 
{ \label{fig:detectors}	 
The pixel side layout is sketched for one of the tiled flip-chip
detectors. The traces and pads are explained in
Section~\ref{sect:flipchip} }  
\end{figure}

Each detector was made with a $4\times4$ array of pixels surrounded by
a 0.55 mm guard ring.  In order to preserve the pixel pitch of 2.5 mm,
the center four pixels were 2.35 mm square and the four corner pixels
were 1.725 mm square. The eight side pixels have a rectangular shape
of 2.35 by 1.725 mm. The gaps between pixels, and between pixels and
guard ring, were $150 \mu$m.  A sketch of the detector pixel layout is
shown in Figure~\ref{fig:detectors}. We also tested a standard
detector from IMARAD.  It was a $20\times20\times4$ mm crystal with an
array of $8\times8$ 1.9 mm indium-contacted pixels. The pixel pitch
was identical to the $4\times4$ array gold-contacted detectors at 2.5
mm.  The cathode plane also used indium.

\subsection{Leakage Current} 
\label{sect:leakage}  

\begin{figure}
\begin{center}
\begin{tabular}{c}
\psfig{figure=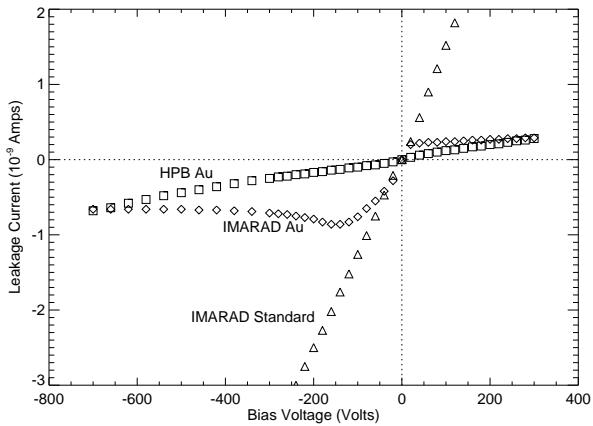,height=6cm} 
\end{tabular}
\end{center}
\caption[iv_curves] 
{ \label{fig:iv_curves}	 
Sample single pixel I-V curves for IMARAD-Au, eV Products-Au, and
standard IMARAD are plotted. The gold contacted IMARAD detector shows 
diode-like curves while the eV Products-Au detector and standard
IMARAD detector show linear curves. }
\end{figure}

We first characterized the detectors by measuring the single-pixel
leakage current as a function of bias voltage.  The reverse bias
resistivity was then calculated at a typical operating voltage of -700
volts. We list the averaged single pixel resistivities in
Table~\ref{tab:resist}. A $3\times3$ array of spring-loaded pogo pins
with conductive rubber tips was used to contact each detector.  By
grounding the exterior eight pins, we were able to isolate the leakage
current to that of one center pixel.  We used a Canberra high voltage
supply and a Keithley 237 as an ammeter.  All measurements were taken
in the dark and at room temperature.  Figure~\ref{fig:iv_curves} shows
the I-V curves taken from a sample pixel of the same geometric size on
the IMARAD-Au and eV Products-Au detectors, and the I-V curve from the
standard IMARAD detector.  Since the pixel sizes and crystal thickness
vary between the IMARAD standard detector and the new gold contacted
detectors, the I-V curves alone do not indicate the difference in
resistivities.  However, one can clearly see the qualitative
difference in the current-voltage relation between the detectors.

\begin{table}
\begin{center}
\begin{tabular}{lccc}
\hline
Detector & Resistivity & $\sigma$ \\
\hline
IMARAD-Au  &  9.9 & 0.28 \\
eV Products-Au  & 10.9 & 0.45 \\
standard IMARAD & 1.0 & 0.2\\
\hline
\end{tabular}
\end{center}
\caption[resist]
{\label{tab:resist} Averaged reverse bias resistivity at -700 V, in
units of $10^{11}$ ohm-cm. The $\sigma$ indicates the empirically
measured rms scatter in the data. }  
\end{table}

The IMARAD standard detector and the eV Products-Au detectors both
exhibited linear I-V curves, while the IMARAD-Au detectors showed a 
diode-like I-V curve.  This indicates that the IMARAD material is
probably n-type since it forms an ohmic contact with a low work
function metal such as indium, while it forms a blocking contact with
high work function metal such as gold.  By using blocking contacts,
the reverse bias resistivity of the IMARAD-Au detector increased by a
factor of ten compared to the standard IMARAD detector.  The IMARAD-Au
resistivity was comparable to that of the eV Products-Au detector at
-700 V.  Furthermore, the flatter slope of the I-V curve shows that
lower leakage current may be possible with the gold contacted IMARAD
detector than with eV Products detector at bias greater than
-700 V.

\subsection{Energy Resolution}
 
\begin{figure}
\begin{center}
\begin{tabular}{c}
\psfig{figure=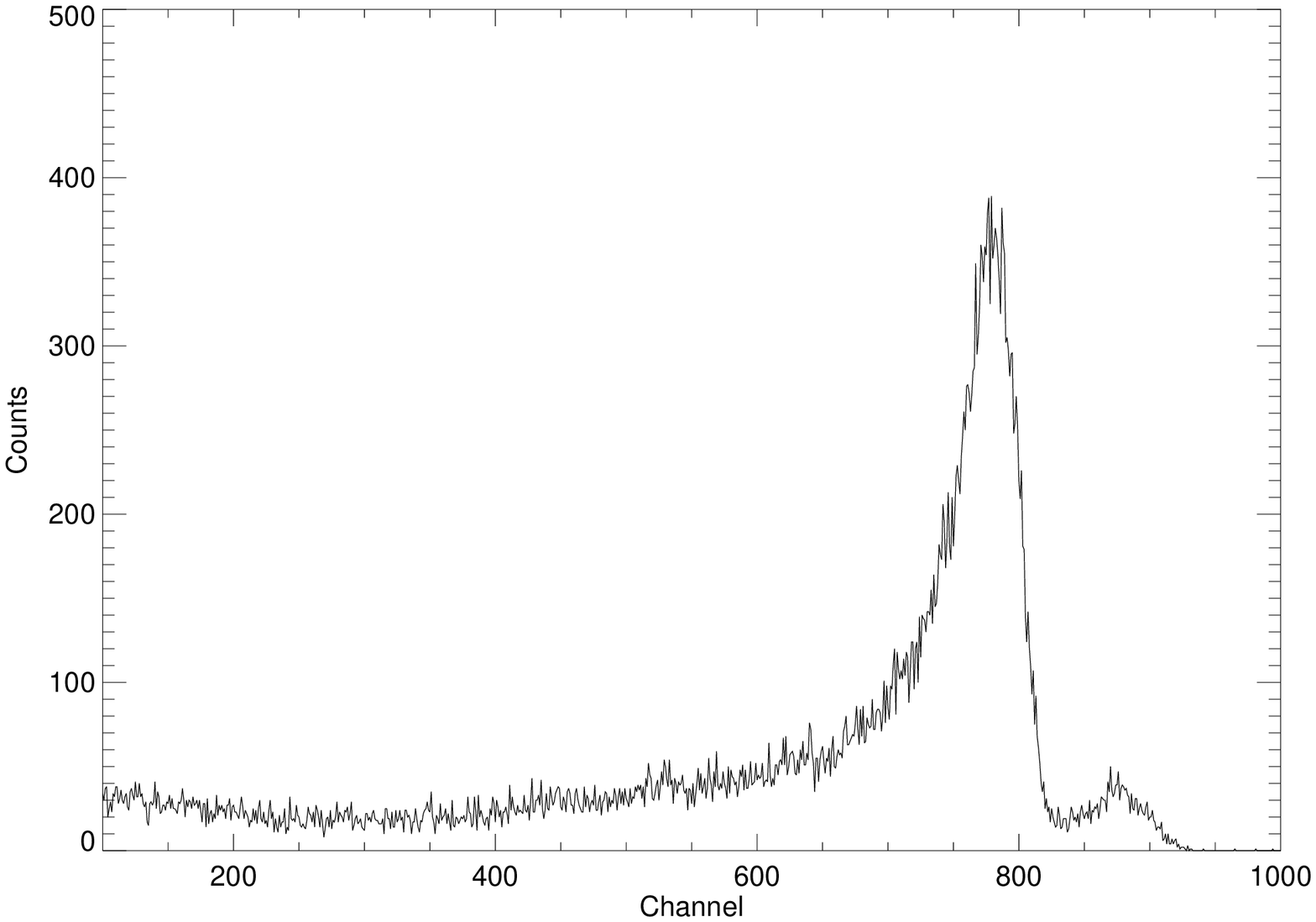,height=5cm} 
\hspace{1cm}
\psfig{figure=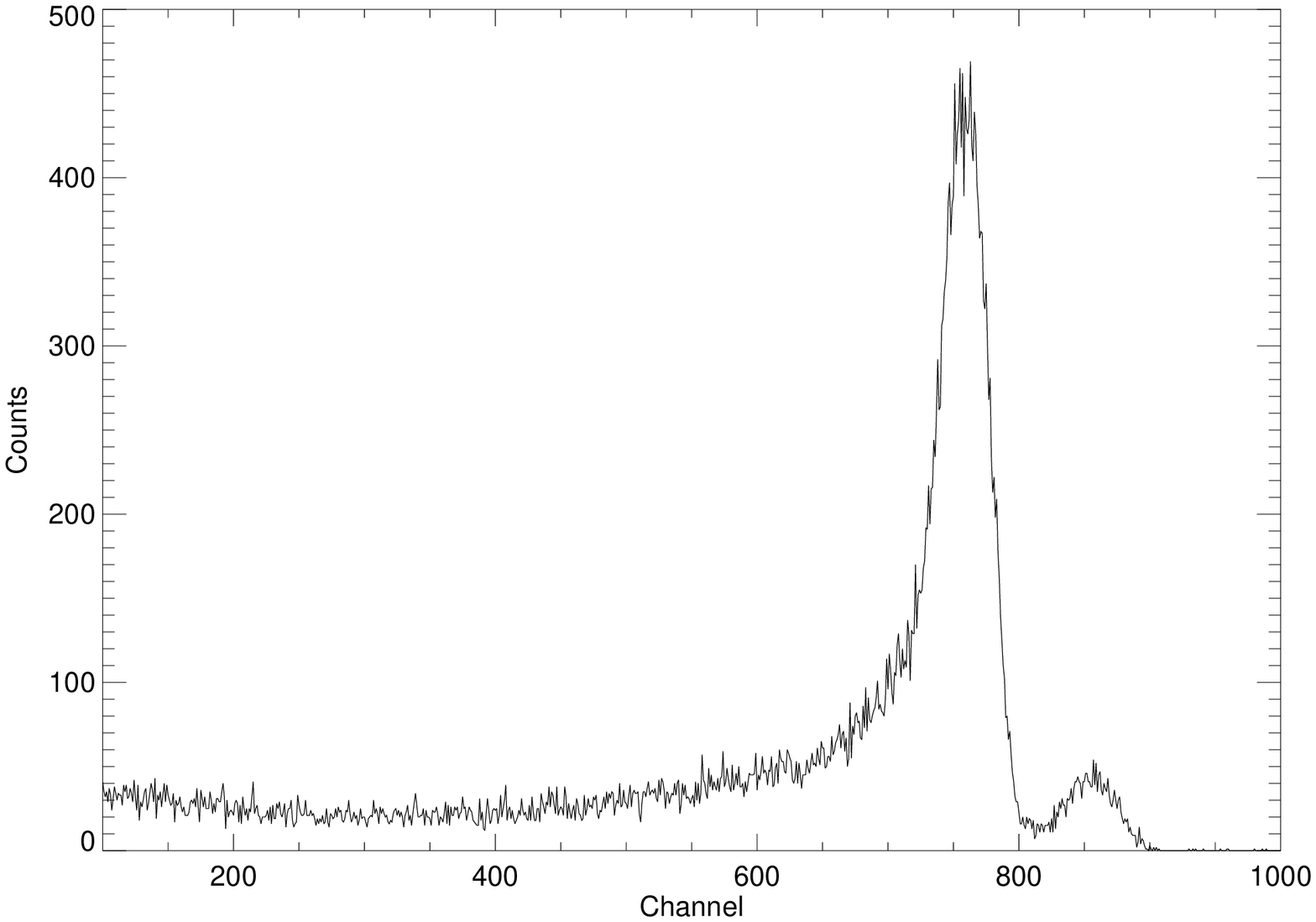,height=5cm} 
\end{tabular}
\end{center}
\caption[Co_spec] 
{ \label{fig:Co_spec}	 
$^{57}$Co spectra taken with eV Products-Au (left) and IMARAD-Au
(right) detectors at -700 V bias.  }   
\end{figure}

The spectral resolution of the detectors, consisting of photopeak
FWHM and photopeak efficiency, was measured using a non-collimated $^{57}$Co
source. The test fixture consisted of an array of spring-loaded pins
contacting the detector. The outside neighbor pixels were shorted
to ground while the center pixel was read out by a low noise eV
Products 550 preamp.  The preamp output was shaped at 1 $\mu$s time
constant and recorded by an MCA.  Typical center pixel $^{57}$Co
spectra are shown in Figure~\ref{fig:Co_spec} for the IMARAD-Au and eV
Products-Au detectors.  The detectors were 
all biased at -700 V. Each spectrum was fit with a Gaussian
photopeak and an exponential low energy tail.  At -700 V bias, the
average center pixel photopeak FWHM energy resolutions (with no
corrections) were $4.2\pm0.6\%$ and $5.5\pm0.6\%$, respectively for
IMARAD-Au and eV Products-Au.  The empirically observed rms scatter in
the FWHM were reported as errors (4 center pixels from 4
detectors). The photopeak efficiency was defined  
as the ratio of the counts in the Gaussian to the total counts in the
range of $+2.35\sigma$ above and $-5\sigma$ below the photopeak
center. The averaged efficiencies were $64.2\pm2.5\%$ and
$76.1\pm2.9\%$ for IMARAD-Au and eV Products-Au detectors, respectively.

\begin{table}
\begin{center}
\begin{tabular}{lcccc}
\hline
Detector & FWHM Energy & $\sigma$ & Photopeak & $\sigma$ \\
  & Resolution ($^{57}$Co) & & Efficiency & \\
\hline
IMARAD-Au & 4.2\% & 0.6\% & 76.1\% & 2.9\%\\
HPB Au & 5.5\% & 0.6\% & 64.2\% & 2.5\%\\
\hline
\end{tabular}
\end{center}
\caption[spectral_fit] 
{ \label{tab:spectral_fit}	 
Average energy resolution and photopeak efficiency from $^{57}$Co
spectra taken with IMARAD-Au and eV Products-Au detectors. The FWHM
values are comparable between detectors, but the IMARAD-Au detectors
have significant improvement in the photopeak efficiency.} 
\end{table}

We found that the energy resolution was slightly better with the
IMARAD-Au detector than with the eV Products-Au detector. This was
surprising given that the pixel leakage currents were similar at the
operating bias of -700 V, and identical readout electronics were used
on both types of detectors.  The difference in the photopeak FWHM was
due to the smaller low energy tail in the IMARAD-Au detector, which
had a significantly better photopeak efficiency than the eV
Products-Au detector.

\begin{figure}
\begin{center}
\begin{tabular}{cc}
\psfig{figure=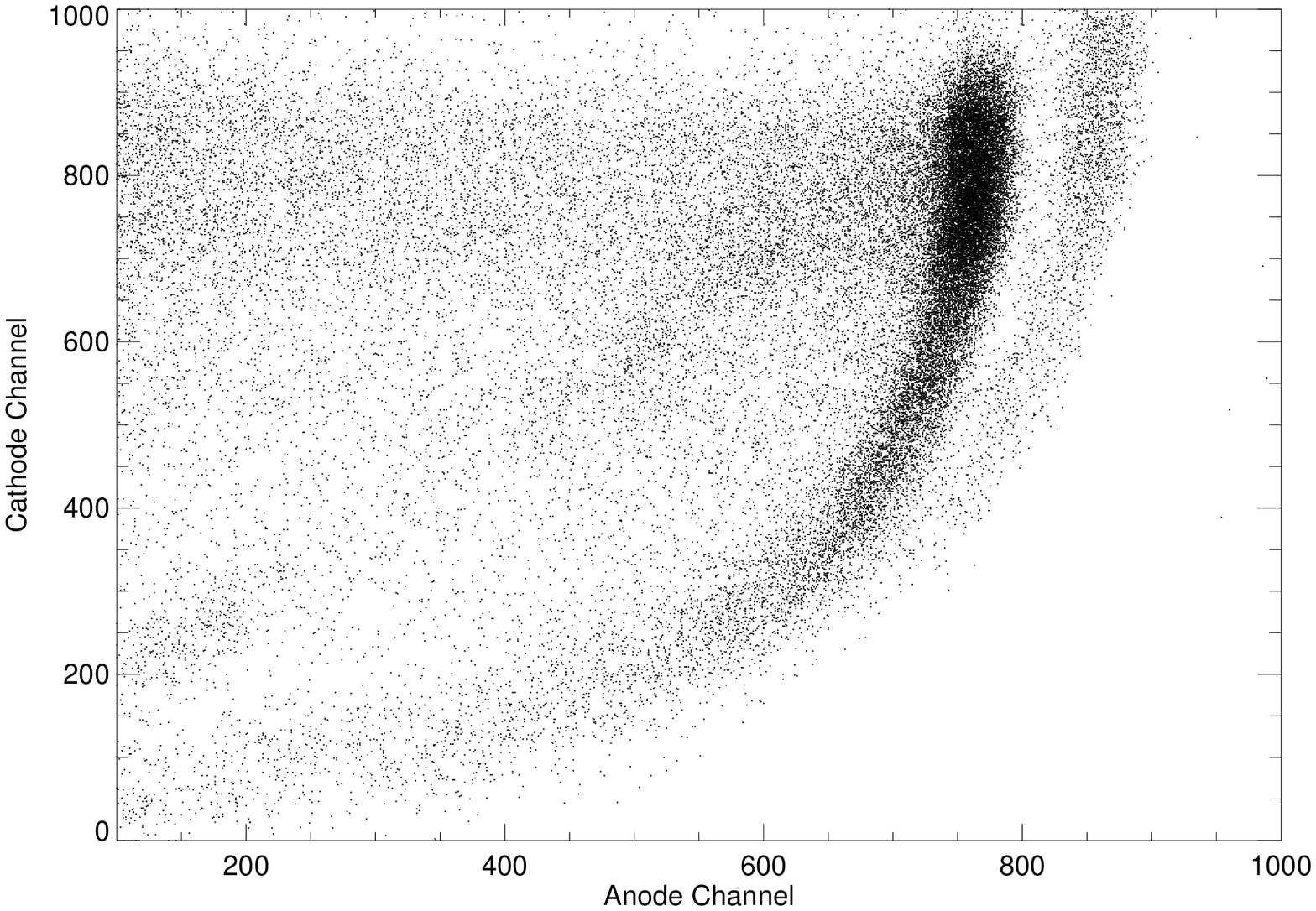,height=5cm} 
\hspace{1cm}
\psfig{figure=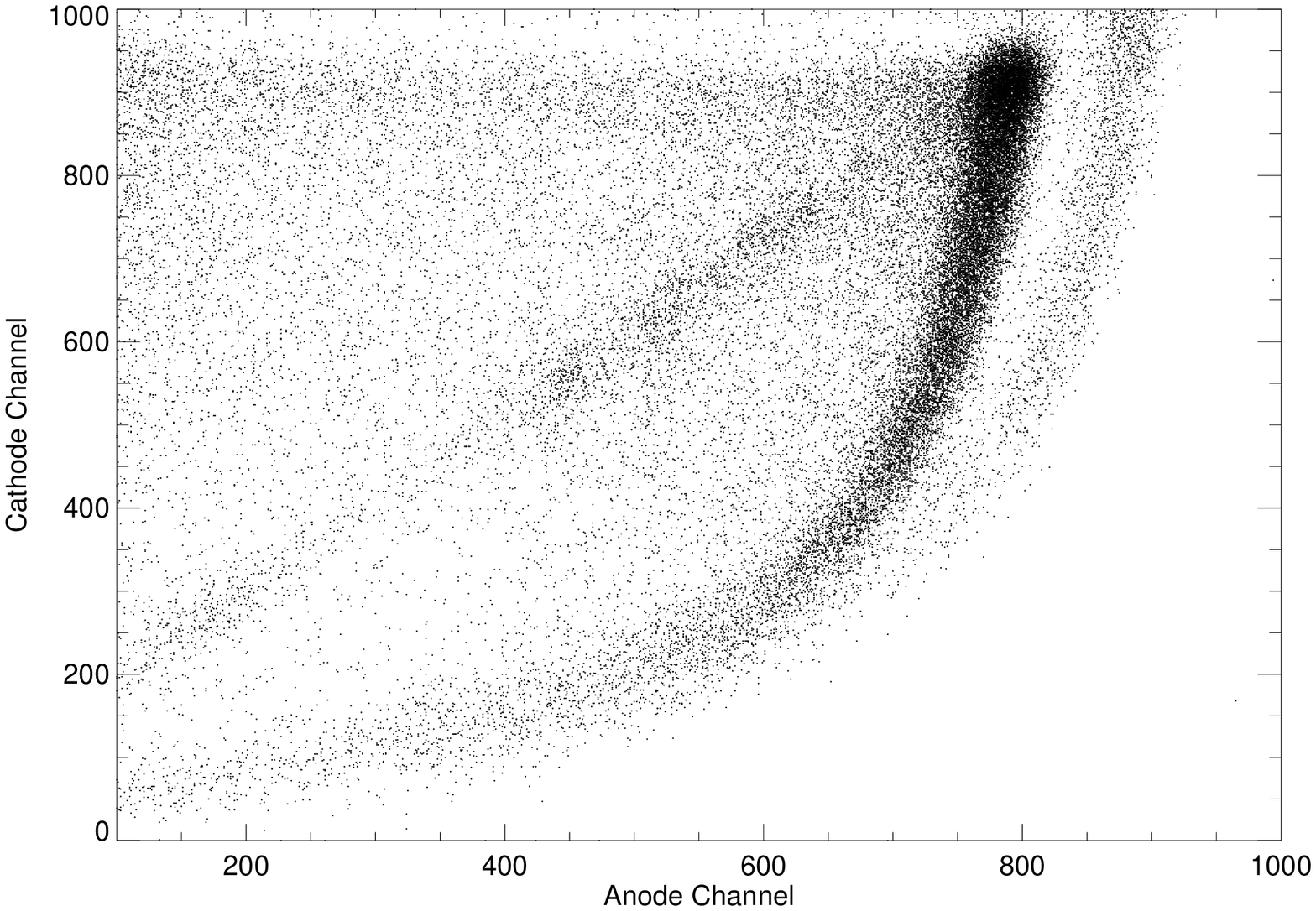,height=5cm} 
\vspace{0.5cm}
\end{tabular}
\begin{tabular}{cc}
\psfig{figure=implot_spie_a.ps,height=5cm} 
\hspace{1cm}
\psfig{figure=evplot_spie_a.ps,height=5cm} 
\end{tabular}
\end{center}
\caption[anode_cathode2] 
{ \label{fig:anode_cathode2}
Comparison of the anode versus cathode
scatter plot between IMARAD-Au and eV Products-Au detectors. The left
column shows the scatter plot and the resulting spectrum for
IMARAD-Au, and the right column shows the same for eV Products-Au. The
photopeak track seen in the scatter plot is typically more vertical
for IMARAD-Au, and its projection on the anode axis results in a less
tailed spectrum.}	
\end{figure}

We investigated this further by measuring the signal induced on the
anode electrode as a function of the X-ray interaction depth. This was
done by digitizing both the anode and the cathode signal for each
X-ray event\cite{shor99}.  While this was not the first use of this
technique to study pixellated devices\cite{schlesinger99}, our study
is the first to compare the performance across two detectors of
identical pixel geometry.  For the anode pixel readout, we used the
same spring loaded pins to contact the center pixel while grounding
the neighbor pixels. The anode signal was then amplified with an eV
Products 550 preamp and shaped with 1$\mu$s time constant. The cathode
signal was also read out by a second eV Products 550 preamp and shaped
with 1$\mu$s time constant.  The shaped anode pulse was used as the
trigger to hold both the anode and cathode pulse heights, which were
then digitized with a 12 bit ADC and recorded by a computer. In
Figure~\ref{fig:anode_cathode2}, the top two panels display a typical
anode versus cathode scatter plot showing the induced signal on the
anode as a function of the interaction depth.  The vertical axis shows the
cathode pulse height and the horizontal axis shows the anode pulse
height.  The set of points on a curved track shows the pulse heights
from 122 keV X-ray photons.  The curvature of the photopeak track is a
function of the weighting potential and the amount of charge trapping.
Since the IMARAD-Au and eV Products-Au detectors have the same pixel
geometry, and thus the same weighting potential, any difference in the
curvature between the detector types must be due to the CZT material
properties.


In Figure~\ref{fig:anode_cathode2}, the bottom two panels show the
projection of the anode versus cathode scatter plot as a spectrum for
each detector. The induced anode signal from the IMARAD-Au detectors
was found to be less dependent on the interaction depth.  This is seen
in the photopeak track which is more vertical for the IMARAD-Au
detectors than for the eV Products detectors. This results in a
spectrum with less tailing and better photopeak efficiency.  The
insensitivity of the anode to the interaction depth is a key feature
of using pixellated devices (small pixel
effect\cite{barrett95}). However, it can also be tuned by 
decreasing the number of electrons moving through the nearly full
thickness of the detector.  By trapping some of the electrons
generated nearest the cathode, which would have otherwise induced the
largest amount of charge on the anode, the amount of induced anode
signal could be made more or less equal for a range of X-ray
interaction depths near the cathode.  An adjustment in the detector
bias voltage can increase the electron drift time, leading to more
trapping.  In our case, the two types of detectors shared the same
pixel geometry and were measured using the same bias voltage. Thus we
speculate that the enhanced photopeak efficiency of the IMARAD-Au
detector must be due to the lower electron mobility in the IMARAD
material.

\subsection{Detector Efficiency} 
\label{sect:Det_eff}

In the past several years, we have experimented with PIN contacts on
CZT to improve the reverse bias resistivity of the
material\cite{narita98}. However, PIN detectors typically suffer from
non-depleted regions where the low electric field limits the movement
of the charge carriers and the X-ray events are lost or degraded.
With our gold-contacted detectors, we compared the active volume of
the IMARAD-Au and eV Products-Au detectors by measuring the detection
efficiency of 122 keV photons. For this test, we used a calibrated
$^{57}$Co beam and recorded the count rate in the center pixels of the
detectors.  The efficiency was defined as the total number of counts
in the spectrum falling between $-5\sigma$ and $+2.35\sigma$ of the
photopeak center. We recorded the efficiency of each center pixel and
averaged them for both detector types.  By dividing the averaged
IMARAD-Au efficiency by the eV Products-Au efficiency, we derived a relative
efficiency between the two detector types of $1.35\pm0.28$. Although
better statistics are needed, our measured relative detector efficiency
is consistent with one. We thus conclude that there is no difference
in the detectors' active volumes, at least up to 122 keV.


\subsection{Temperature Sensitivity} 

CZT detectors in space missions may be subject to some variation in
temperature unless active thermal control systems are employed.
Thus, we are interested in the performance 
of these detectors in the temperature range of $-20^{\circ}$ to
$+20^{\circ}$ C. To study the 
temperature sensitivity of IMARAD detectors, we placed the IMARAD-Au
detector and the IMARAD standard detectors in a thermal chamber,
cooled them to $-20^{\circ}$ Celsius, and recorded $^{241}$Am and $^{57}$Co
spectra. The test fixture used the same spring-loaded pin array and
the low noise preamp for measuring energy resolution at room
temperature. The shaping time was fixed at $1\mu$s. The
entire electronic readout system was cooled along with the detectors,
and the temperature of the detector was measured by a thermistor
attached to the test fixture.  To minimize the thermal shock to the
CZT, the cooling chamber was controlled such that the temperature slew
of the detector fixture was about one degree per minute.

\begin{figure}
\begin{center}
\begin{tabular}{cc}
\psfig{figure=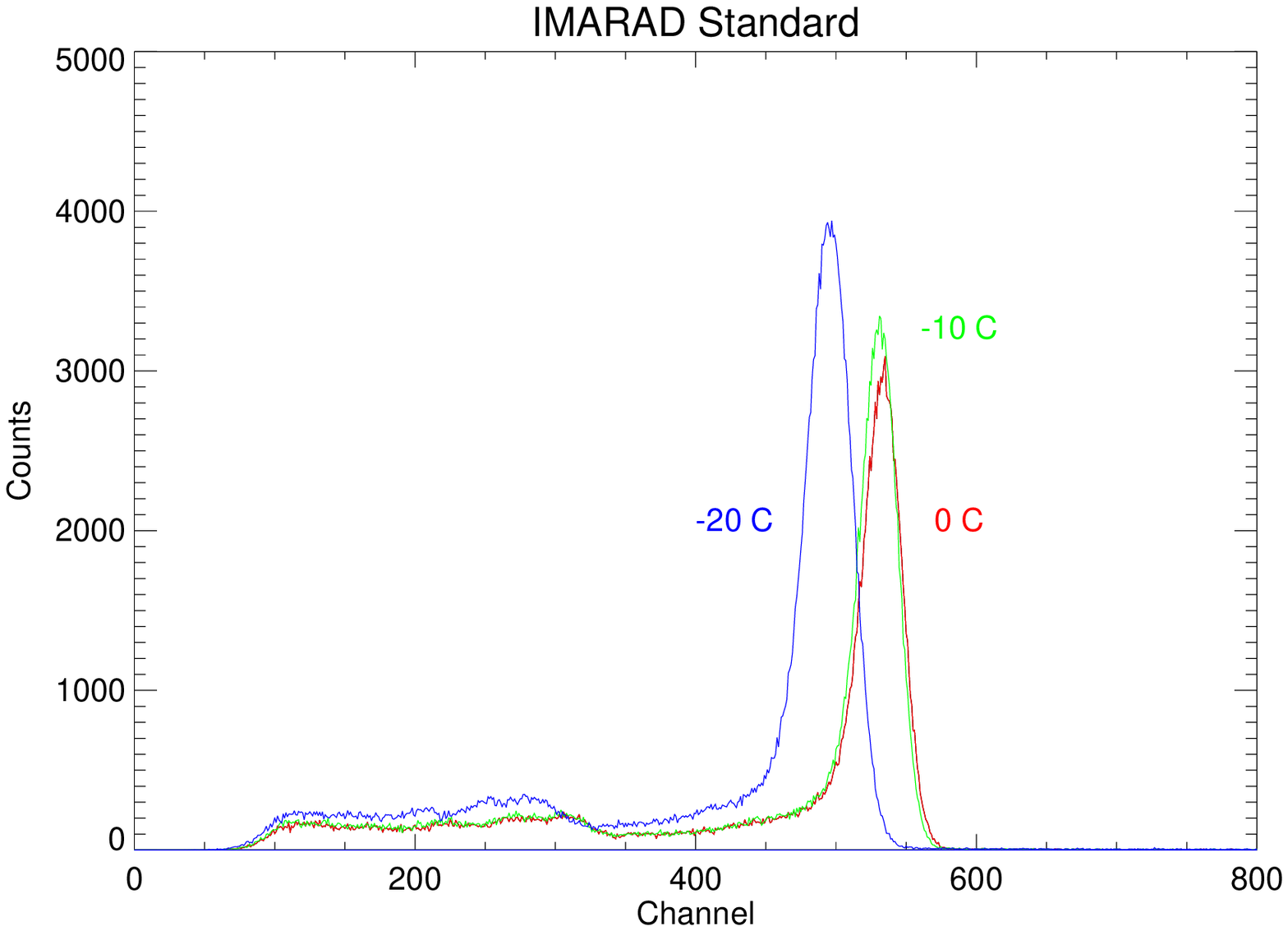,height=5cm} 
\psfig{figure=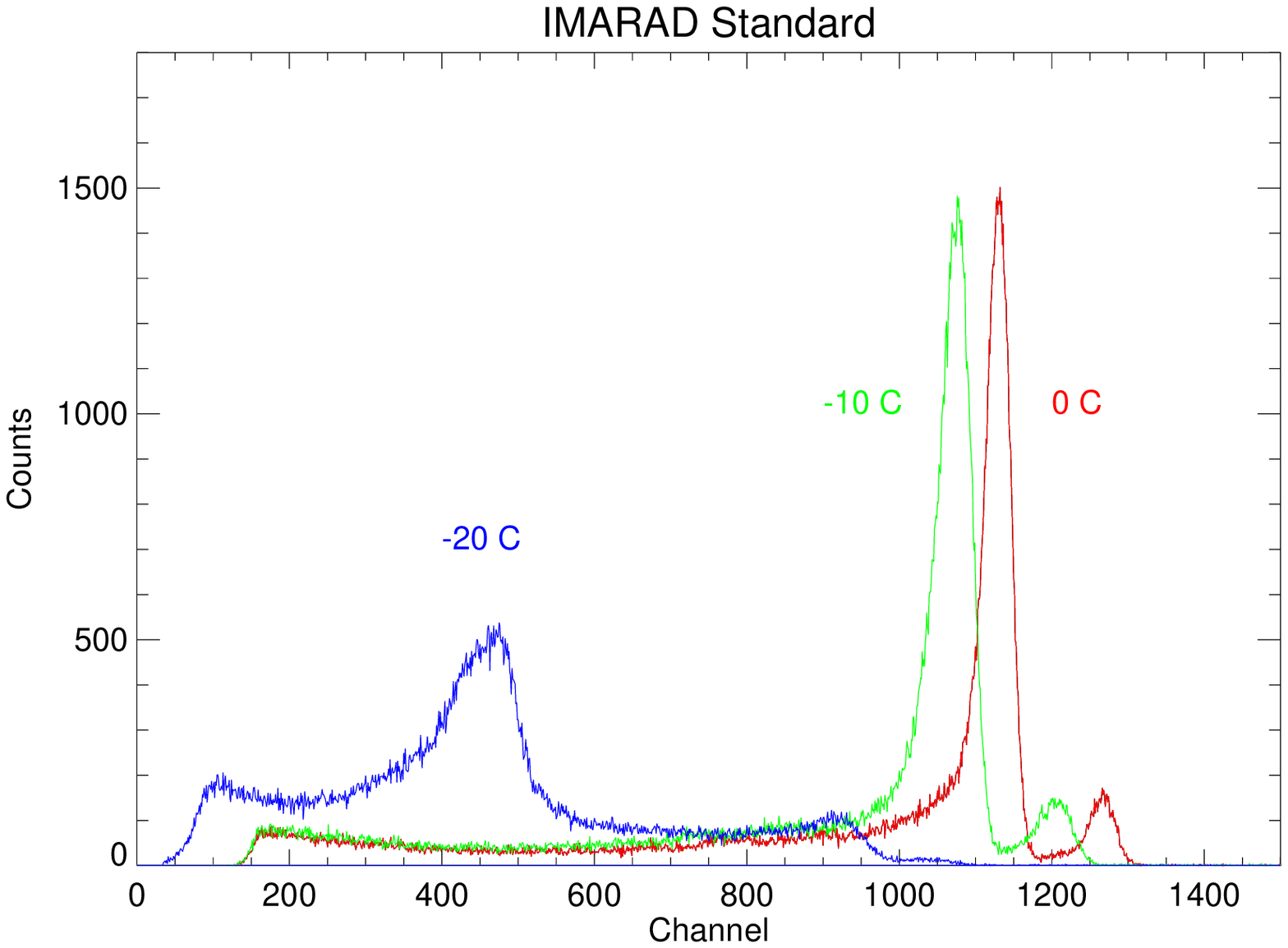,height=5cm} 
\end{tabular}
\begin{tabular}{cc}
\psfig{figure=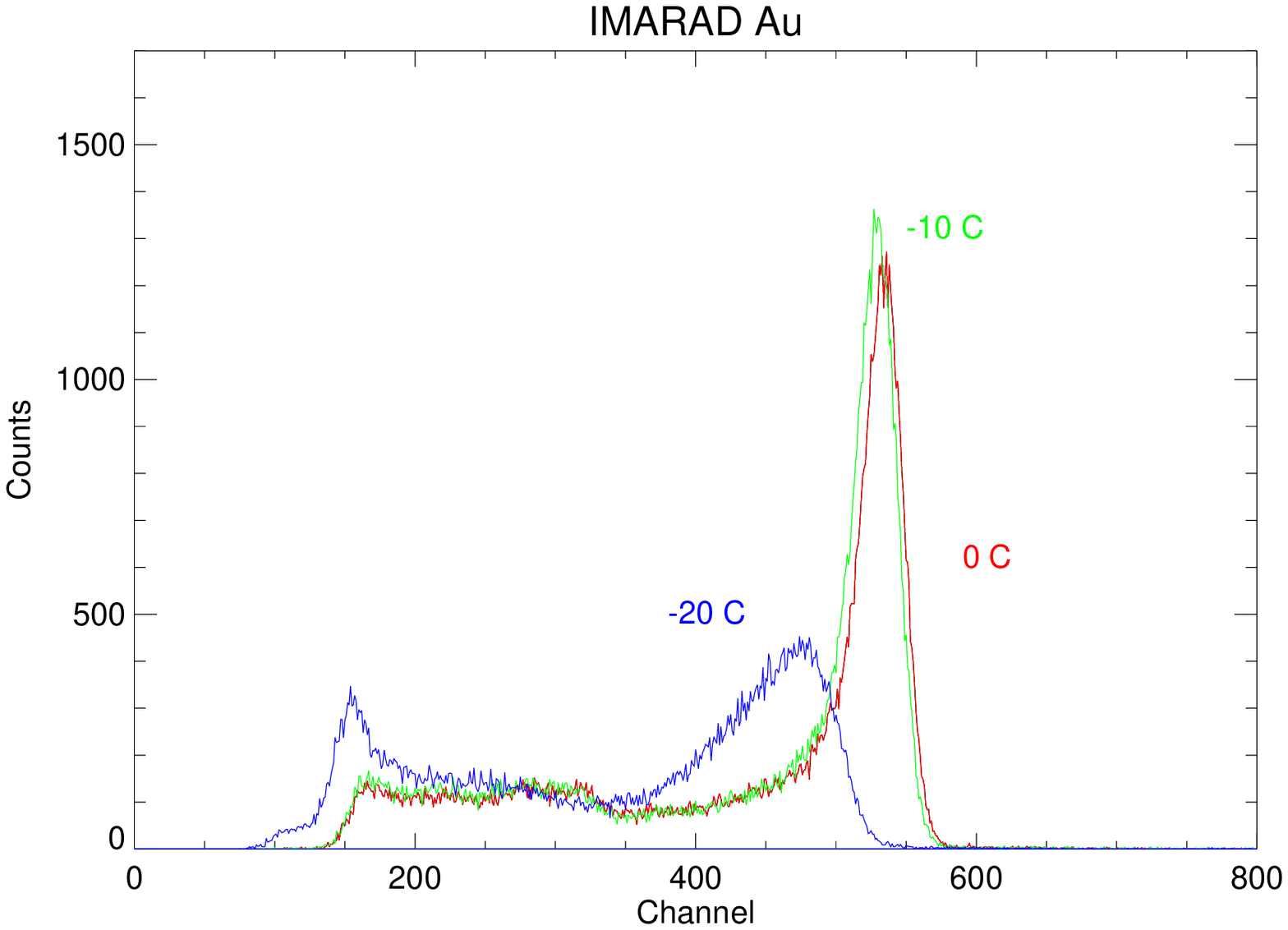,height=5cm} 
\psfig{figure=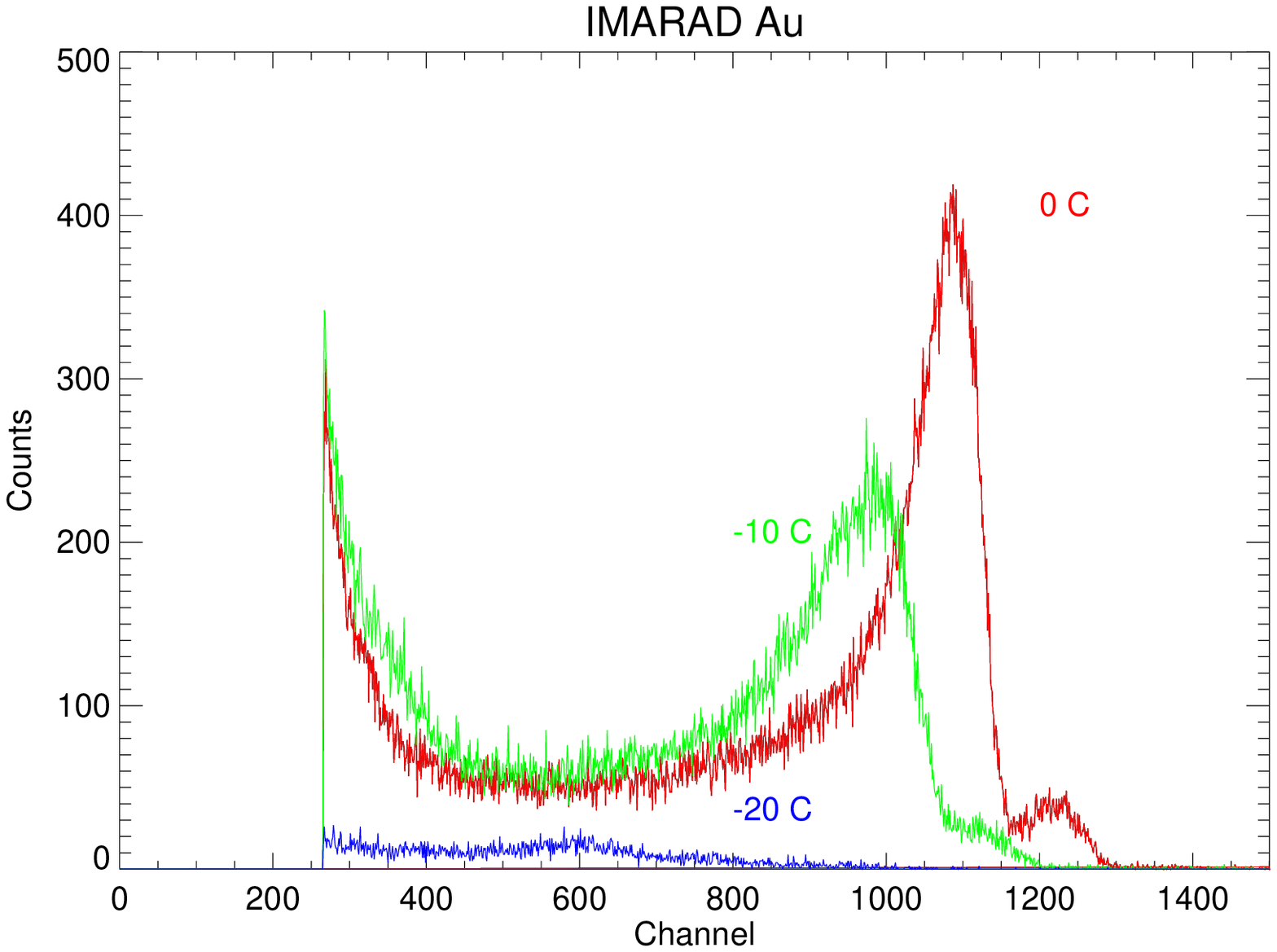,height=5cm} 
\end{tabular}
\end{center}
\caption[lowt_spec] 
{ \label{fig:lowt_spec}	
$^{241}$Am (left column) and $^{57}$Co (right column) spectra taken with
IMARAD standard and IMARAD-Au detectors at 0, -10, and -20$^{\circ}$. 
} 
\end{figure}

Each spectrum was integrated for one minute while the temperature of
the detector stayed constant within $\pm1^{\circ}$ C. The detectors
were all biased at -700 V.  Figure~\ref{fig:lowt_spec} shows a mosaic
of spectra taken with IMARAD standard and IMARAD-Au detectors, at $0$,
-10, and $-20^{\circ}$Celsius.  For the standard IMARAD detector, the
photopeak gain decreased with lower temperature, but interestingly the 
$^{241}$Am photopeak count rate increased from $-10^{\circ}$ to
$-20^{\circ}$.  This was not the case for the $^{57}$Co spectrum where
the 122 keV photopeak degraded below $-10^{\circ}$.  The photopeak
degradation was even more apparent for the IMARAD-Au detector.  The
$^{241}$Am photopeak was severely degraded below $-10^{\circ}$, and
the 122 keV photopeak had almost disappeared at $-20^{\circ}$ in a
manner similar to the polarization effects found in CdTe detectors. 
Although we are currently investigating the low temperature behavior
in more detail, we speculate that these effects are caused by an increase
in the electron detrapping time.  Such an effect would result in a
poor spectrum due to incomplete charge collection and ballistic
deficit for the $1\mu$s shaping time used.

\subsection{Flip-chip Detector} 
\label{sect:flipchip}

\begin{figure}
\begin{center}
\begin{tabular}{c}
\psfig{figure=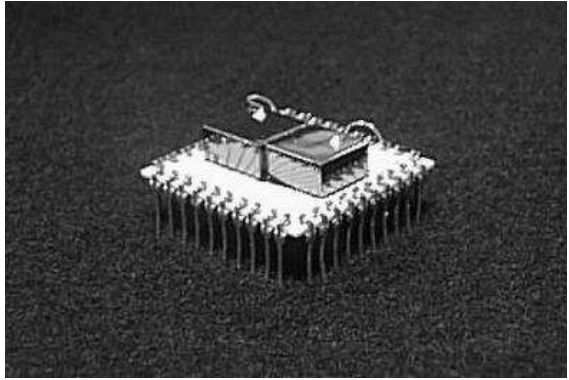,height=5cm} 
\end{tabular}
\end{center}
\caption[flipchip_pic] 
{ \label{fig:flipchip_pic}	
Tiled flipchip detector array on a readout board. The wires connect
the high voltage pads to the cathode plane.} 
\end{figure}

As a preliminary step to building a tiled large area detector, we
fabricated and tested two tiled flip-chip detector modules using one
IMARAD-Au detector and one eV Products-Au detector on each module.  Each
detector module required a thin Alumina substrate board with an array
of $4\times8$ gold pixel pads at 2.5 mm pitch. Wire bond gold studs
were placed in the center of each pad. Additional pads were placed for
the guard ring and the high voltage connections.  Fanout traces
connected the internal pads to the edge pads, where clip-pins were
soldered. A sketch of the Alumina board is shown in
Figure~\ref{fig:detectors}. The two crystals were visually aligned and
bonded in a tiled 
manner onto the pixel pads using conductive epoxy. After a low
temperature curing cycle, a non-conductive epoxy underfill was used for
additional adhesion.  A picture of the detector module is shown in
Figure~\ref{fig:flipchip_pic}. The entire process was done at HyComp,
Inc. in Marlborough, MA. 

Due to unforeseen problems in the fabrication process, three pixels on
the first flip-chip module IMARAD-Au detector developed shorts. Even
more IMARAD-Au pixels were shorted on the second flip-chip module,
rendering it unusable. Although the cause is not clear, we speculate
that it was due to surface oxidation, preferentially on IMARAD CZT,
from the underfill epoxy which produced partially conductive
paths. The behavior of the remaining channels was normal. The detector
module is readout by a 32-channel VA/TA ASIC made by IDE AS. One
detector module and the support electronics will be flown on a balloon
payload in September 2000 to measure the hard X-ray background
spectrum at 125,000 feet.  More detailed performance and simulation
results from the first flip-chip module are given in Bloser et
al. 2000\cite{bloser00}

\section{CONCLUSIONS} 
\label{sect:conclusion}  

In order to realize the goal of developing a large area hard X-ray
detector for astronomy, the yield of spectroscopic grade large area
detectors must be increased.  Current crystal growth processes using
the HPB technique result in CZT with high resistivity but relatively
low yield, whereas the HB growth process has high yield but low
resistivity. To compensate for the low resistivity in HB CZT, we have
fabricated and tested pixellated detectors using gold blocking
contacts on IMARAD CZT.

By using simple evaporated gold electrodes, the operating bias
resistivity was increased by an order of magnitude compared to the
IMARAD detectors made without the blocking contacts. The resistivity
of the gold-contacted IMARAD detector was comparable with that of the
higher resistivity HPB CZT detectors when both detectors were biased
at -700 V.  The spectra taken with the IMARAD-Au detector showed
excellent photopeak efficiency, and this was probably due to higher
level of electron trapping in the IMARAD CZT compared to the HPB CZT
from eV Products. A similar effect might also result from a
detector with a low bias region near the cathode, but we found no
difference in the detector efficiency between IMARAD-Au and eV
Products-Au detectors using a calibrated $^{57}$Co source.  We did
find however, that the IMARAD detectors performed poorly when they
were cooled to below $-10^{\circ}$ Celsius.  We speculate that this is
a further increase in the high level of electron trapping in the
material.  Finally, we fabricated and tested a tiled flip-chip
detector for use in our hard X-ray balloon-borne experiment. The
detector, comprising one IMARAD-Au and one eV Products-Au detector,
will be flown in September 2000, and the results will be given in a
future paper.

\acknowledgments     

We thank Dr. Uri El-Hanany of IMARAD Imaging Systems and Kevin Parnham
at eV Products for supplying us with CZT samples.  We thank Kanai Shah
and Paul Bennett at RMD Inc. for their assistance in detector
fabrication.  Finally, 
we thank George Riley at HyComp, Inc. for his assistance in the
flipchip processing.   This work was supported in
part by NASA grants NAG5-5103 and NAG5-5209.


  \bibliography{paper}   
  \bibliographystyle{spiebib}   
 
  \end{document}